\documentstyle[12pt]{article}
\begin{document}
\begin{titlepage}
\begin{flushright}
IPNO/TH 95-60\\
\end{flushright}
\vfill
\centerline{\Large\bf Systematic 1/M Expansion for Spin 3/2 Particles}
\centerline{\Large\bf in Baryon Chiral Perturbation Theory}
\vfill
\centerline{\large Thomas R. Hemmert and Barry R.
Holstein\footnote{Research supported in part by the National 
Science Foundation}}
\vspace*{0.5cm} 
\centerline{Department of Physics and Astronomy}
\centerline{University of Massachusetts}
\centerline{Amherst, MA 01003, U.S.A.}
\vspace*{1cm}
\centerline{\large Joachim Kambor$\footnote{Laboratoire de Recherche des 
Universit\'es Paris XI et Paris VI, associ\'e au CNRS.}$ }
\vspace*{0.5cm}
\centerline{Division de Physique Th\'eorique} 
\centerline{Institut de Physique Nucl\'eaire} 
\centerline{F-91406 Orsay Cedex, France}
\vfill
\begin{abstract}
Starting from a relativistic formulation of the pion-nucleon-delta system,
the most general structure of 1/M corrections for a  heavy baryon chiral 
lagrangian including spin 3/2 resonances is given.
The heavy components of relativistic nucleon and delta fields are 
integrated out and their contributions to the next-to-leading order lagrangians
are constructed explicitly.
The effective theory obtained admits a systematic expansion
in terms of soft momenta, the pion mass $m_\pi$ and the delta-nucleon mass 
difference $\Delta$.
As an application, we consider neutral pion photoproduction at threshold 
to third order in this small scale expansion.
\end{abstract}
\vfill
\end{titlepage}
\newpage

\section{Introduction}

Chiral symmetry provides important restrictions on 
the interactions of pions, nucleons and photons
\cite{CA}. The consequences are most conveniently summarized by 
the use of
an effective field theory, valid in the low energy regime. This simultaneous 
expansion in small momenta and light quark masses is known as Chiral 
Perturbation Theory (ChPT) \cite{Pag75,Wei79,BSW85,GL85a,GSS88}. 
Unlike in the sector of Goldstone bosons,
the mass of the nucleon is large and nonvanishing in the chiral limit. 
Nevertheless, a consistent chiral power counting, known as Heavy 
Baryon Chiral Perturbation Theory, HBChPT, can be maintained by performing
also a systematic $1/M$-expansion, $M$ being the mass of a baryon
\cite{JM91a,JM91}.\footnote{Such techniques have been developed and used
extensively in the parallel case of heavy quark physics---{\it cf.} 
\cite{AB,EM}.}  In principle, any observable of the pion-nucleon system 
can be 
calculated to a given order in the chiral expansion---the price to be paid 
is the introduction of new low energy coupling constants which are not fixed 
by the symmetry requirement alone.  Any parameter-free prediction of HBChPT
is, however, also a 
prediction of low energy QCD.  Many of these resulting ``low energy theorems'' 
(LET) have been
discussed in the recent literature \cite{BKM95}.

The spin $3/2$ delta resonances play a special role 
in the pion-nucleon system, since the mass difference $\Delta=M_\Delta-M_N$
is not large compared to the typical low energy scale $m_\pi$ and because
the $\pi N \Delta$-coupling constant is anomalously large.  
The more conventional 
version of HBChPT takes into account the effect of the delta (and of other 
resonances) only through contributions to the coupling constants of higher 
order operators in the chiral expansion. 
This approach is in particular well suited to derive low energy theorems. 
A concern, however, is that, in the physical 
world of nonvanishing quark masses, the perturbation series might converge 
slowly due to the presence of large coupling constants driven by small 
denominators---{\it
i.e.} by terms proportional to $1/\Delta$. An alternative approach to HBChPT 
includes the delta degrees of 
freedom explicitly \cite{JM91,BSS93}. In addition to solving 
the problems mentioned above, this technique has 
the advantage that the range of applicability can in principle be
extended into the delta-region.

In this letter we sketch explicitly 
the steps necessary for a systematic low energy 
expansion in the presence of the spin 3/2 delta resonance.  A full
presentation will appear shortly.\cite{HHK95}  We begin with 
a covariant formulation of an effective theory of the $\pi N \Delta$-system. 
The heavy degrees of freedom are identified and integrated out via
a systematic $1/M$-expansion. We arrive at an effective field theory of 
nonrelativistic nucleons and deltas coupled to pions and external sources. 
The theory is manifestly Lorentz invariant and admits a low energy expansion 
in terms of small momenta $q$, the pion mass $m_\pi$ 
and the delta-nucleon mass difference $\Delta$, which we collectively
denote by the symbol $\epsilon$.\footnote{We shall continue to refer to
the conventional chiral treatment, wherein only $q$ and $m_\pi$ are
taken as small quantities, as an expansion in the generic small
quantity $p$.}  Of course, the procedure described in the next section
is not unique---the general methods of such heavy mass expansions
have been given previously \cite{AB,EM,LM92}.  
However, it represents a useful starting
point for the evaluation of higher order effects .  Indeed, the 1/M
corrections derived in this manner have a simple physical
interpretation---exchange of the heavy degrees of freedom---and the
formalism can straightforwardly be extended to deal with higher order
terms in the $\epsilon$ expansion, as will be discussed in
\cite{HHK95}.  Furthermore, it is straightforward to treat other resonances
than spin 3/2 along the same lines.

As a simple application of our formalism we shall consider neutral 
pion photoproduction at threshold . A one-loop calculation within the framework
of HBChPT has produced a LET for the electric dipole amplitude
$E_{0^+}$ as an expansion in powers of $\mu=m_\pi/M_N$
\cite{BGKM91,BKKM92}
\begin{equation}
E_{0+}^{\pi^0 p}(s_{\rm thr})=-{e g_{\pi NN} \over 8\pi M_N} \left[ \mu
- {1\over 2}(3+\kappa_p) \; \mu^2 - \left({M_N\over 4 F_\pi}\right)^2 \; \mu^2
+{\cal O}(\mu^3) \right]
\label{LET}
\end{equation}
Here $\kappa_p$ is the anomalous magnetic moment of the proton while 
$g_{\pi NN}$ is the strong pion-nucleon coupling constant. 
Recently an ${\cal O}(p^4)$ 
calculation has been given\cite{BKM94} which reconciles the 
theoretical prediction with experiment \cite{Fuchs95}. However, each term in 
this expansion is large with alternating sign, thus making the 
convergence particularly slow.  Below we calculate the correction of order 
$\epsilon^3$ to Eq. (\ref{LET}), which arises due to a 1/M corrected vertex
of the $\Delta$(1232) resonance.
However, first we set our formalism.

\section{1/M-expansion for spin 3/2 resonances}

Consider the lagrangian for a relativistic spin 3/2 field $\Psi_\mu$ coupled 
in a chirally invariant manner to the Goldstone bosons
\begin{equation}
{\cal L}_{3/2}=\bar\Psi^\alpha O^A_{\alpha\mu} \Lambda^{\mu\nu} O^A_{\nu\beta}
\Psi^\beta
\label{eq:la3/2}
\end{equation}
with
\begin{equation}
O^A_{\alpha\mu}=g_{\alpha\mu}+{1\over 2} A \gamma_\alpha \gamma_\mu . 
\end{equation}
Following Pascalutsa \cite{Pas94} we have factored out the dependence on 
the unphysical free parameter A by use of the projection operator 
$O^A_{\alpha\mu}$. Then defining the physical spin 3/2 field as 
\begin{equation}
\psi_\mu (x)=O^A_{\mu\nu} \Psi^\nu (x) \label{eq:redef}
\end{equation}
we see that Eq. (\ref{eq:la3/2}) is manifestly invariant under point 
transformations
\begin{eqnarray}
\Psi_{\mu}(x) & \rightarrow & \Psi_{\mu}(x)+\lambda \gamma_{\mu}\gamma_{\nu}
                              \Psi^{\nu}(x) \nonumber \\
A             & \rightarrow & \frac{A-2 \lambda}{1+4\lambda}
\end{eqnarray}
as required by general considerations \cite{Nath}. 

To leading order in the derivative expansion, the relativistic spin 3/2 
lagrangian with field-redefinition Eq.(\ref{eq:redef}) then takes the form
\footnote{To take into account the isospin 3/2 property of $\Delta$(1232) we
supply the Rarita-Schwinger spinor with an additional isospin index $i$, 
subject to the subsidiary condition $\tau^i \; \psi_{\mu}^i (x) = 0$.}
\begin{equation}
{\cal L}_{\Delta}=\bar{\psi}^{\mu}_i \; \Lambda_{\mu\nu}^{ij} \; \psi^{\nu}_j 
\end{equation}
with
\begin{eqnarray}
\Lambda_{\mu\nu}^{ij} & = & - \mbox{\Large [} ( i \not\!\!{D}^{ij} - M_{\Delta}
                            \; \delta^{ij} ) g_{\mu \nu} - \frac{1}{4} 
                            \gamma_{\mu} \gamma^{\lambda} ( i \not\!\!{D}^{ij}
                            - M_{\Delta} \; \delta^{ij}) \gamma_{\lambda} 
                            \gamma_{\nu} \nonumber \\
                      &   & \; \; \; + \frac{g_{1}}{2} g_{\mu \nu} 
                            \not\!{u}^{ij} 
                            \gamma_{5} + \frac{g_{2}}{2} ( \gamma_{\mu} 
                            u_{\nu}^{ij} + u_{\mu}^{ij} \gamma_{\nu} ) 
                            \gamma_{5} + \frac{g_{3}}{2} \gamma_{\mu} 
                            \not\!{u}^{ij} \gamma_{5} \gamma_{\nu} 
                            \mbox{\Large ]} \label{eq:Lambda}
\end{eqnarray}
Following the conventions of SU(2) HBChPT in the spin 1/2 sector 
\cite{GSS88,BKM95}, we have defined the following structures:
\begin{eqnarray}
D_{\mu}^{ij} \; \psi^{\nu}_j & = & \left( \partial_{\mu} \; \delta^{ij} + 
                                   \Gamma_{\mu}^{ij} \right) \psi^{\nu}_j 
                                   \nonumber \\
\Gamma_{\mu}^{ij} & = & \Gamma_\mu \; \delta^{ij} - \frac{i}{2} \; 
                        \epsilon^{ijk} \; Tr [ \tau^k \; \Gamma_\mu ]
                        \nonumber \\
\Gamma_{\mu}       & = & \frac{1}{2} \left[ u^{\dag} , \partial_{\mu} u \right]
                         - \frac{i}{2} u^{\dag} ( {\bf v}_{\mu} + 
                         {\bf a}_{\mu} ) u - \frac{i}{2} u ( {\bf v}_{\mu} - 
                         {\bf a}_{\mu} ) u^{\dag} 
                         \nonumber \\
u_\mu^{ij}         &=&  u_\mu\delta^{ij}-i\epsilon^{ijk}w_\mu^k \nonumber\\
w_\mu^i            &=& \frac{1}{2} Tr 
                         \left[ \tau^i u_\mu \right] \nonumber \\
u_{\mu}            & = & i u^{\dagger} \nabla_{\mu} U u^{\dagger}\nonumber\\
\nabla_{\mu} U     & = & \partial_{\mu} U - i ( {\bf v}_{\mu} + {\bf a}_{\mu} 
                         ) U + i U ( {\bf v}_{\mu} - {\bf a}_{\mu} ) 
                         \nonumber \\
U                  & = & u^{2} = exp \left( \frac{i}{F_{\pi}} {\bf \vec{\tau} 
                         \cdot \vec{\pi} } \right) 
\label{eq:definitions}
\end{eqnarray}
${\bf v}_\mu$, ${\bf a}_\mu$ denote external vector, axial-vector fields and 
are the only external sources possible at this  order. 
The first two pieces in Eq. (\ref{eq:Lambda}) are the kinetic and
mass terms of a free spin 3/2 lagrangian \cite{Pas94}. The remaining terms
constitute the most general chiral invariant couplings to pions. Note that 
aside from the conventional $\pi\Delta\Delta$ coupling constant $g_1$
we have included two additional pion-couplings characterized by 
$g_2$, $g_3$ which contribute only if at least one of the spin 3/2
fields is off mass shell. 

The next step consists of identifying the ``light'' and ``heavy'' 
degrees of freedom of the spin 3/2 fields, respectively. The procedure is 
analogous to the case of spin 1/2 fields, as pioneered in the case of heavy 
quark effective theory \cite{AB,EM} and later applied to spin 1/2
HBChPT \cite{BKKM92}. For the case of spin 3/2 particles the problem is 
technically somewhat more challenging due to the off-shell spin 1/2 degrees 
of freedom associated with the Rarita-Schwinger field\footnote{The formalism 
for the case of heavy systems of arbitrary spin was given by Falk in 
\cite{AB}.}. In order to separate the spin 3/2 from the spin 1/2 components 
it is convenient to introduce a complete set of orthonormal spin projection 
operators for fields with {\it fixed velocity} $v_\mu$
\begin{eqnarray}
P^{3/2}_{(33)\mu \nu} & = & g_{\mu \nu} - \frac{1}{3} \gamma_{\mu} \gamma_{
                               \nu} - \frac{1}{3} \left( \not\!{v} \gamma_{\mu} 
                               v_{\nu} + v_{\mu} \gamma_{\nu}\not\!{v} \right) 
                               \nonumber \\
P^{1/2}_{(11)\mu \nu} & = & \frac{1}{3} \gamma_{\mu} \gamma_{\nu} - v_{\mu} 
                               v_{\nu} + \frac{1}{3} \left( \not\!{v}  
                               \gamma_{\mu} v_{\nu} + v_{\mu} \gamma_{\nu} 
                               \not\!{v} \right)  
                               \nonumber \\
P^{1/2}_{(22)\mu \nu} & = & v_{\mu} v_{\nu}\nonumber\\
P^{1/2}_{(12)\mu\nu}&=&{1\over \sqrt{3}}(v_\mu v_\nu -\not\!{v}
v_\nu\gamma_\mu)\nonumber\\
P^{1/2}_{(21)\mu\nu}&=&{1\over \sqrt{3}}(\not\!{v} v_\mu\gamma_\nu-v_\mu v_\nu)
\label{eq:project}
\end{eqnarray}
which satisfy
\begin{eqnarray}
P^{3/2}_{(33)\mu \nu} + P^{1/2}_{(11)\mu \nu} + P^{1/2}_{(22)\mu \nu} 
                       & = &  g_{\mu \nu} \nonumber \\
P^{I}_{(ij)\mu \nu} \; P^{J, \nu \delta}_{(kl)} 
                       & = & \delta^{I J} \; \delta_{jk} \; P^{J, 
                             \delta}_{(il)\mu} .
\label{eq:orthogonality}
\end{eqnarray}  
The four-velocity $v_\mu$ is related to the four-momentum $p_\mu$ of the 
spin 3/2 particle by
\begin{equation}
p_\mu=M v_\mu+k_\mu ,
\label{eq:residual}
\end{equation}
where $M$ is a baryon mass scale and $k_\mu$ is taken to be a residual soft 
momentum. We now employ the familiar projection operators of the heavy mass 
formalism 
\begin{equation}
P_{v}^{\pm} = \frac{1}{2} \left( 1 \pm \not\!{v} \right).
\end{equation}
and introduce heavy baryon fields for our spin 3/2 particles in order to 
eliminate the dependence on the large mass $M_\Delta$ in Eq.(\ref{eq:Lambda}). In analogy to the spin 1/2 case we identify the ``light'' spin 3/2 degree of 
freedom via
\begin{equation}
T_{\mu}^i (x) \equiv  P_{v}^{+} \; P^{3/2}_{(33)\mu\nu} \; \psi^{\nu}_i (x) 
                    \; \mbox{exp}(i M v \cdot x) \label{eq:T}
\end{equation}
whereas the remaining components 
\begin{equation}
G_{\mu}^i (x)  =  \left( g_{\mu\nu}-P_{v}^{+} \; P^{3/2}_{(33)\mu\nu}\right) 
                   \psi^{\nu}_i (x) \; \mbox{exp}(i M v \cdot x) 
\label{eq:G}
\end{equation}
can be shown to be ``heavy'' \cite{HHK95} and are integrated 
out. We note in particular that $G_{\mu}$ includes both spin 1/2 {\it and} 
spin 3/2 components.  Of course, the 
virtual effects of the heavy degrees of freedom $G_{\mu}$ are nevertheless
accounted for in the heavy baryon formalism, they show up as higher order
1/M corrected vertices involving the remaining (on-shell) spin 3/2 fields 
$T_\mu$, as we will show below.
We also note that the $T_{\mu}$ degrees of freedom satisfy the constraints
\begin{equation}
v_\mu T^{\mu}_i=\gamma_\mu T^{\mu}_i=0 \label{eq:subsidiary}
\end{equation}
and correspond to the SU(2) version of the decuplet field introduced in 
ref.\cite{JM91}.

We now perform a systematic $1/M$-expansion, following an approach developed
by Mannel et al. in HQET \cite{EM}, which was later applied to spin 1/2 
HBChPT by Bernard et al. \cite{BKKM92}. Since we are interested in the 
interactions of nucleons with the spin 3/2 resonance, we must treat both 
fields simultaneously. We therefore write the most general lagrangian 
involving relativistic spin 1/2 ($\psi_N$) and spin 3/2 ($\psi_\mu$) fields as 
\begin{equation}
{\cal L}={\cal L}_N + {\cal L}_\Delta + \left( {\cal L}_{\Delta N} + h.c. 
\right) \label{eq:4L}
\end{equation}
with ${\cal L}_\Delta$ given in Eq.(\ref{eq:Lambda}) and
\begin{eqnarray}
{\cal L}_N &=& \bar\psi_N \left(i\not\!\!{D}-M_N+{g_A\over 2}\not\!{u} \gamma_5
\right) \psi_N + ... \nonumber \\
{\cal L}_{\Delta N} &=& g_{\pi N \Delta} \; \bar{\psi}^{\mu}_{i} \left(
g_{\mu\nu}+z \gamma_\mu \gamma_\nu \right) w^{\nu}_{i} \psi_N + ...
\end{eqnarray}
where the dots denote higher order counterterm contributions and $z$ 
corresponds to the leading-order pion-nucleon-delta off-shell coupling 
constant.

Rewriting the lagrangians of Eq.(\ref{eq:4L}) in terms of the spin 3/2 
heavy baryon components $T_{\mu}$ and $G_{\mu}$, and the corresponding
``light'' and ``heavy'' spin 1/2 components $N$, $h$, defined as
\begin{eqnarray}
N(x)&=& P_v^+ \; \psi_N \; {\rm exp} (i M v\cdot x) \nonumber \\
h(x)&=& P_v^- \; \psi_N \;  {\rm exp} (i M v\cdot x), 
\label{heavyN}
\end{eqnarray}
we find the general heavy baryon lagrangians
\begin{eqnarray}
L_{N} &=& \bar{N} {\cal A}_{N} N + \left( \bar{h} {\cal B}_{N} N + h.c. \right)
          - \bar{h} {\cal C}_{N} h \nonumber \\
L_{\Delta N} &=& \bar{T} {\cal A}_{\Delta N} N + \bar{G} {\cal B}_{\Delta N} N +
                 \bar{h} {\cal D}_{N \Delta} T + \bar h {\cal C}_{N \Delta} G +
                  h.c. \nonumber\\
L_{\Delta}&=&\bar{T} {\cal A}_{\Delta} T + \left( \bar{G} {\cal B}_{\Delta} T +
             h.c. \right) - \bar{G} {\cal C}_{\Delta} G.
\label{Lgeneral}
\end{eqnarray}
Note that we have used the {\it same} mass $M$ in the definition of heavy 
delta and nucleon fields respectively. This is necessary in order that 
all exponential factors drop out in Eq. (\ref{Lgeneral}). The matrices 
${\cal A}_{N}$, ${\cal B}_{N}$, ...,  ${\cal C}_{\Delta}$ admit a small 
energy scale expansion of the form
\begin{equation}
{\cal A}_\Delta = {\cal A}_\Delta^{(1)}+{\cal A}_\Delta^{(2)}+ ... ,
\end{equation}
where ${\cal A}_\Delta^{(n)}$ is of order $\epsilon^n$.  As emphasized in the 
introduction, we denote by $\epsilon$ small quantities of order $p$, like 
$m_\pi$ or soft momenta, {\it as well as} the mass difference 
$\Delta=M_\Delta-M_N$. This mass difference is distinct from the pion mass 
in the sense that it stays finite in the chiral limit. However, in the 
physical world, $\Delta$ and $m_\pi$ are of the same magnitude. We therefore 
adhere to a simultaneous expansion in both quantities.  It is only through 
this small scale expansion that we obtain a systematic low energy expansion 
of the $\pi N \Delta$-system. 

To make this more explicit, consider the leading order matrices 
${\cal A}_{N}^{(1)}, \; {\cal A}_{\Delta N}^{(1)}$ and 
${\cal A}_{\Delta}^{(1)}$. Choosing the heavy baryon mass parameter $M=M_N$, 
we obtain
\begin{eqnarray}
{\cal A}_N^{(1)} &=& i(v\cdot D)+g_A (S\cdot u) \nonumber \\
{\cal A}_{\Delta N}^{(1)}  &=& g_{\pi N\Delta} w_{\mu}^i \nonumber\\
{\cal A}_{\Delta}^{(1)} &=& - \left[ i \; v \cdot D^{ij} - \Delta \; 
\delta^{ij} + g_{1} \; S \cdot u^{ij} \right] \; g_{\mu \nu}  
\label{eq:llead}
\end{eqnarray}
where $S_\mu$ denotes the Pauli-Lubanski spin vector. One can easily see
from Eq.(\ref{eq:llead}) that our formalism produces the exact SU(2) analogues
of the spin 1/2 \cite{JM91a} and spin 3/2 \cite{JM91} lagrangians of Jenkins 
and Manohar. Furthermore, as expected, the {\cal O}($\epsilon$) heavy baryon 
lagrangians Eq.(\ref{eq:llead}) are free of the off-shell couplings $z , g_2, 
g_{3}$ 
\footnote{This point has also been emphasized in recent work by Napsuciale and
Lucio \cite{NL96}.}  In our formalism off-shell couplings 
will only start contributing at ${\cal O}
(\epsilon^2$) via ${\cal B}$ and ${\cal D}$ matrices. Explicit expressions for
the expansions of ${\cal B}_\Delta$, ${\cal C}_\Delta$ {\it etc}. will not be 
displayed here but can be found in ref. \cite{HHK95}. 

From Eq.(\ref{eq:llead}) we determine the SU(2) HBChPT propagator for
the delta field:
\begin{equation}
i \; S_{\mu\nu}^{3/2} ( v \cdot k ) = \frac{- i}{v \cdot k - \Delta + i0^+}
                                      \; P^{3/2}_{\mu\nu} \; \xi^{ij}_{I=3/2} 
\label{eq:propagator}
\end{equation}
where $P^{3/2}_{\mu\nu}$ is a spin 3/2 projector \cite{JM91} 
\begin{equation}
P^{3/2}_{\mu\nu} = P_{v}^{+} \; P^{3/2}_{(33)\mu\nu} \; P_{v}^{+} = g_{\mu\nu}
- v_\mu v_\nu + \frac{4}{3} \; S_\mu S_\nu 
\end{equation}
and 
\begin{equation}
\xi^{ij}_{I=3/2} = \delta^{ij} - \frac{1}{3} \; \tau^i \tau^j
\end{equation}
denotes an isospin 3/2 projector. From Eq.(\ref{eq:propagator}) one can 
see that the delta propagator counts as $\epsilon^{-1}$ in our expansion 
scheme.

The final step is again in analogy to the heavy mass formalism for spin 1/2
systems. Shifting variables and completing the square, we obtain the effective
action
\begin{equation}
S_{\rm eff}= \int d^4x \left\{ \bar T \tilde {\cal A}_{\Delta} T
+\bar N \tilde {\cal A}_{N} N
+\left[ \bar T \tilde {\cal A}_{\Delta N} N + h.c.\right] \right\}
\label{Seff}
\end{equation}
with
\begin{eqnarray}
\tilde {\cal A}_\Delta &=& {\cal A}_\Delta 
+ \gamma_0 \tilde {\cal D}_{N \Delta}^\dagger \gamma_0 \tilde {\cal C}_N^{-1} 
\tilde {\cal D}_{N \Delta}
+ \gamma_0 {\cal B}_\Delta^\dagger \gamma_0 {\cal C}_\Delta^{-1} {\cal B}_\Delta
\nonumber \\
\tilde {\cal A}_N &=& {\cal A}_N 
+ \gamma_0 \tilde {\cal B}_{N}^\dagger \gamma_0 \tilde {\cal C}_N^{-1} 
\tilde {\cal B}_{N}
+ \gamma_0 {\cal B}_{\Delta N}^\dagger \gamma_0 {\cal C}_\Delta^{-1} 
{\cal B}_{\Delta N}
\nonumber \\
\tilde {\cal A}_{\Delta N} &=& {\cal A}_{\Delta N} 
+ \gamma_0 \tilde {\cal D}_{N \Delta}^\dagger \gamma_0 \tilde {\cal C}_N^{-1} 
\tilde {\cal B}_{N} 
+ \gamma_0 {\cal B}_\Delta^\dagger \gamma_0 {\cal C}_\Delta^{-1} 
{\cal B}_{\Delta N} 
\label{eq:Atilde}
\end{eqnarray}
and 
\begin{eqnarray}
\tilde {\cal C}_N&=& {\cal C}_N-{\cal C}_{N \Delta}
{\cal C}_{\Delta}^{-1} \gamma_0 {\cal C}_{N \Delta}^\dagger \gamma_0 
\nonumber \\
\tilde {\cal B}_N&=& {\cal B}_N+{\cal C}_{N \Delta }
{\cal C}_{\Delta}^{-1} {\cal B}_{\Delta N} \nonumber \\
\tilde {\cal D}_{N\Delta}&=& {\cal D}_{N \Delta}
+ {\cal C}_{N \Delta} {\cal C}_{\Delta}^{-1} {\cal B}_{\Delta} .
\label{BCtilde}
\end{eqnarray}
Eq.(\ref{eq:Atilde}) represents the master formula of our treatment of a
coupled spin 1/2 - spin 3/2 system in HBChPT. All 1/M corrected vertices
can be directly obtained by calculating the appropriate matrices
${\cal A}, {\cal B}, {\cal C}, {\cal D}$ to any order desired. The new terms 
proportional to ${\cal C}_\Delta^{-1}$ and ${\cal C}_N^{-1}$
are given entirely in terms of coupling constants of the lagrangian for 
relativistic fields. This guarantees reparameterization\footnote{Imposing the 
conditions of reparameterization invariance on the leading order lagrangians 
Eq.(\ref{eq:llead}) as developed by Luke and Manohar for the case of HQET 
\cite{LM92} provides an alternative way to derive the 1/M corrected vertices.}
and Lorentz invariance \cite{LM92,EM95}. Furthermore, all such terms are 
$1/M$ suppressed. The effects of the heavy degrees of freedom (both spin 3/2 
and 1/2) thus show up only at order $\epsilon^2$. Note also that 
the effective $NN$-, $N\Delta$- and $\Delta\Delta$-interactions all contain
contributions from both heavy $N$- and $\Delta$-exchange respectively.  

In the above formalism, it is understood that at each order 
one must also include the most 
general counterterm lagrangian consistent with chiral symmetry, Lorentz 
invariance, and the discrete symmetries P and C. As should be
clear, it is crucial to write this counterterm lagrangian in terms of 
relativistic fields. The choice of variables Eqs.(\ref{eq:T},\ref{eq:G}) and
Eq.(\ref{heavyN}) yields
then automatically the contributions to matrices ${\cal A}$, 
${\cal B}$, ${\cal C}$ and ${\cal D}$. Only these objects have a well 
defined small scale expansion---since $S_{\rm eff}$ is written 
entirely in terms of heavy baryon fields, derivatives count as order 
$\epsilon$, quark masses as order $\epsilon^2$ etc. In order to calculate a 
given process to order $\epsilon^n$, it thus suffices to construct matrices 
${\cal A}$ to the same order, $\epsilon^n$, ${\cal B}$ and ${\cal D}$ to 
order $\epsilon^{n-1}$, and ${\cal C}$, $\tilde {\cal C}$ to order 
$\epsilon^{n-2}$. Note that since the propagator of the $T_\mu$ field counts 
as order $\epsilon^{-1}$,
one-particle reducible diagrams as in Fig. 1 have also to be considered. 
\footnote{This could be avoided by imposing systematically the equations of 
motion for asymptotic states, as done in ref. \cite{EM95} for the nucleon 
sector.} 
Finally one has to add all loop-graphs contributing at the order one is
working. The relevant diagrams can be found by straightforward power 
counting in $\epsilon$.

\section{1/M corrections to threshold $\pi^0$ photoproduction to 
{\cal O}($\epsilon^3$)}

As an elementary example of the use of this formalism, consider neutral
pion photoproduction\footnote{A complete HBChPT analysis of the effects of 
$\Delta$(1232)
on S and P-wave multipoles in pion photoproduction is in preparation.}. 
A phenomenological analysis of the
influence of $\Delta$(1232) in this process was given in \cite{DM88}.
At threshold in the small scale expansion
as described above, to order $\epsilon^3$, this amounts to calculating all 
1-loop graphs with vertex insertions from $\tilde {\cal A}^{(1)}$ as well as
all tree graphs with vertices derived from up to and including 
$\tilde {\cal A}^{(3)}$ to the process $\gamma p \rightarrow p \pi^0$. The 
electric dipole amplitude is then related to the cross section in the center
of mass frame through \cite{DT92}
\begin{equation}
(E_{0+})^2={|{\bf k}| \over |{\bf q}|} 
{d\sigma \over d\Omega}|_{|{\bf q}|\rightarrow 0} ,
\end{equation}
where ${\bf k}$ and ${\bf q}$ are the photon and pion three-momenta, 
respectively.

It is most convenient to break up the calculation into one-particle irreducible
(1PI) diagrams. Possible loop graphs involving the leading order vertices
of Eq. (\ref{eq:llead}) start at order $\epsilon^3$ in our counting. However, 
one can check explicitly that, aside from the well-known triangle graph 
contribution \cite{BGKM91,BKKM92}, {\em at threshold} there exist no 
other loop effects to $E_{0+}^{\pi^0 p}$ involving $\Delta$(1232) to this 
order in the $\epsilon$-expansion.

The photoproduction amplitude is then given by the diagrams of
Fig. 1 a)-c). Due to the structure of ${\cal A}_{N}^{(1)}$ and 
${\cal A}_{\Delta N}^{(1)}$, several simplifications appear: at threshold the 
nonvanishing 1PI subgraphs of Fig. 1 are at least $O(\epsilon^2)$. For the 
nucleon-nucleon transition this is well known. For the nucleon-delta vertex, 
the situation is similar. To leading order,
the $\gamma N\Delta$ vertex does not exist, the $\gamma\pi N\Delta$ coupling
vanishes for the neutral pion, and the $\pi N\Delta$ vertex is proportional to
$q^\mu$, which, when contracted with projection operator 
$P^{3/2}_{\mu\nu}$ associated with the delta-propagator, 
vanishes at threshold. Moreover, 1PI-vertices without pions or photons
attached, also only begin at $O(\epsilon^2)$; this is the reason why no tree 
diagrams with more than a single propagator need be considered. 
Thus the 1PI one-photon and one-pion vertices are needed to $O(\epsilon^2)$, 
while the 1PI $\pi\gamma$ vertices are to be calculated to $O(\epsilon^3)$.

Analyzing Eq.(\ref{eq:Atilde}) we evaluate the following structures for our 
calculation of $E_{0+}^{\pi^0 p}$ at {\cal O}($\epsilon^3$):
\begin{center}
\begin{tabular}{l|l}
vertex & lagrangian \\ \hline
O($\epsilon^2$) $\gamma NN$ & ${\cal A}_N^{(2)}$ and
 $\gamma_0 {\cal B}_N^{(1) \; \dagger} \gamma_0 
{\cal C}_N^{(0) \; -1} {\cal B}_N^{(1)}$ \\
O($\epsilon^2$) $\pi^0 NN$ & 
$\gamma_0 {\cal B}_N^{(1) \; \dagger} \gamma_0 {\cal C}_N^{(0) \; 
-1} {\cal B}_N^{(1)}$ \\
O($\epsilon^2$) $\gamma\Delta N$ & ${\cal A}_{\Delta N}^{(2)}$ \\ 
O($\epsilon^2$) $\pi^0 \Delta N$ & 
 $\gamma_0 {\cal B}_\Delta^{(1) \; \dagger} \gamma_0 
{\cal C}_\Delta^{(0) \; -1} {\cal B}_{\Delta N}^{(1)}$ \\
O($\epsilon^2$) $\pi^0 \gamma NN$ & 
$\gamma_0 {\cal B}_N^{(2) \; \dagger} \gamma_0 
{\cal C}_N^{(0) \; -1} {\cal B}_N^{(1)} + h.c.$ \\
O($\epsilon^3$) $\pi^0 \gamma NN$ & vanishes at threshold 
\end{tabular} \newline
\end{center}
Vertices which do not involve spin 3/2 particles can be taken from 
\cite{BKM95,EM95}. Summing up all the nucleon-only contributions, including 
the triangle graphs, we recover the LET Eq.(\ref{LET}),
as expected. We now proceed to analyse the effects of 1/M corrected vertices
involving $\Delta$(1232).

Due to the fact that the photoexcitations of $\Delta$(1232) only start with 
the M1 transition, there is no $\gamma N\Delta$ interaction at 
{\cal O}($\epsilon$). Consequently, there is also no 1/M corrected vertex at 
{\cal O}($\epsilon^2$). However, the well-known relativistic counterterm 
lagrangian  
\cite{HHK95,DMW91}
\begin{equation}
{\cal L}_{c.t.}^{\gamma N\Delta}={i b_1\over 2 M_N} \bar\psi^{\mu}_i 
\left( g_{\mu\nu} + y \gamma_\mu \gamma_\nu \right) \gamma_\rho \gamma_5
\; \frac{1}{2} \; Tr \left[ f_{+}^{\rho\nu} \; \tau^i \right] \psi_N,
\end{equation}
provides a large part of the M1 transition strength and leads to the heavy 
baryon structure
\begin{equation}
{\cal A}_{\Delta N}^{(2)}={i b_1 \over M_N} S_\nu \; \frac{1}{2} \; Tr \left[ 
f_+^{\nu\lambda} \; \tau^i \right],
\label{gammavertex}
\end{equation}
which we use in the diagrams of Fig. 1b.

The leading order $\pi N\Delta$ vertex does not contribute at threshold, but 
its {\em $1/M$ corrected structure} can provide a contribution to the s-waves !
Multiplying out the relevant matrices, we find:
\begin{equation}
\left( \gamma_0 {\cal B}^{(1) \; \dagger}_\Delta \gamma_0 
{\cal C}_\Delta^{(0) \; -1} {\cal B}_{\Delta N}^{(1)} 
\right)_{\Delta N}= \frac{-i}{M_N} \; g_{\pi N\Delta} \; D_{\mu}^{ij}
(v\cdot w^j).
\label{pivertex}
\end{equation}

These two vertices then lead to an {\cal O}($\epsilon^3$) contribution of 
$\Delta$(1232) to the process $\gamma p \rightarrow \pi^0 p$ at threshold,
given by the diagrams in Fig. 1b:
\begin{equation}
E_{0+}^\Delta={e \over {8\pi}}{4 b_1 g_{\pi N\Delta} \over {9F_\pi M_{N}^2}} 
{m_\pi^3 \over {m_\pi+\Delta}}.
\label{E0+new}
\end{equation}
This new contribution of Eq. (\ref{E0+new}) is distinct in the
following sense---in the chiral limit, it scales like 
$m_\pi^3$, {\it i.e.} the corresponding 
photoproduction amplitude is of order $p^4$. The LET Eq.(\ref{LET}) is
therefore not violated by this term. There are {\it many} other terms which
are of O($p^4$) \cite{BKM94}, but Eq.(\ref{E0+new}) is the {\it only} term
which is of order $\epsilon^3$ due to the 1/M corrections.  
In the physical world of finite pion mass, $E_{0+}^{\Delta}$ is in principle 
of the same order of magnitude as the $p^3$ effects. Moreover, it has the 
opposite sign to the large $p^3$ terms in the LET for $E_{0+}$. 

In order to give a numerical estimate for $E_{0+}$ to O($\epsilon^3$), we add 
Eq. (\ref{E0+new}) and Eq. (\ref{LET}). Utilizing $b_1=-2.30 \pm 0.35 , 
g_{\pi N\Delta}=1.5\pm 0.2$ \cite{DMW91}, we find
\begin{eqnarray}
E_{0+}^{\pi^0 p} = - \; {e g_{\pi N} \over 8\pi M_N} \; \mu 
\times\left\{ \begin{array}{ll}
+ 1 & \mbox{{\cal O}($\epsilon^2$) Kroll-Ruderman term} \\
- 0.97  & \mbox{{\cal O}($\epsilon^3$) $N\pi$ loop graphs} \\
- 0.35  & \mbox{{\cal O}($\epsilon^3$) Nucleon Born terms}\\
+ 0.07  & \mbox{{\cal O}($\epsilon^3$) 1/M corrected Delta
       terms} \end{array}\right. \nonumber 
\end{eqnarray}
\begin{equation}
\rightarrow E_{0+}^{\pi^0 p} \approx 0.8\times 10^{-3}/m_\pi . 
\end{equation}
Comparing with the number extracted from the most recent experiment 
\cite{Fuchs95}, $E_{0+}=(-1.31\pm 0.08)\times 10^{-3}/m_{\pi^+}$, which is in 
agreement with a chiral O($p^4$) calculation \cite{BKM94}, we conclude that 
it is mandatory
to calculate the $E_{0+}$ multipole to O($\epsilon^4$). As the O($p^4$) 
calculation shows, $N\pi$ loop graphs at the next order cancel to a large 
extent the big contribution from the triangle graphs. It will be interesting 
to see how big the $\Delta$(1232) effects are at 
that order. Work in this direction is under progress.

To conclude, we have presented a systematic low energy expansion of HBChPT
including spin 3/2 resonances. As an application, we have considered $\pi^0$
photoproduction at threshold and have found the leading contribution
of $\Delta$(1232) to $E_{0+}^{\pi^0 p}$ to be of order
$m_\pi^3/(m_\pi+\Delta)$. Many other processes associated
with the $\pi N\Delta$-system can be treated with the formalism described 
here.

\newpage
\quad\\
\vspace{4in}
\quad\\
\noindent Figure 1: Nucleon pole (a), delta pole (b), and contact (c) 
diagrams contributing to pion photoproduction in the heavy baryon approach.
\end{document}